\documentstyle[epsfig,cite]{mn2e}  

\begin{document}

\title[A radio jet in the prototypical symbiotic star Z And?]
    {A radio jet in the prototypical symbiotic star Z And?}
\author[Brocksopp et al.]
    {C.~Brocksopp$^{1,2}$\thanks{email: cb4@mssl.ucl.ac.uk}, J.L. Sokoloski$^3$, C. Kaiser$^4$, A.M. Richards$^5$, T.W.B. Muxlow$^5$,
\newauthor
 N. Seymour$^6$\\
\\
$^1$ Mullard Space Science Laboratory, University College London, Dorking, Surrey RH5 6NT\\
$^2$ Astrophysical Research Institute, Liverpool John Moores University, Twelve Quays House, Egerton Wharf, Birkenhead CH41 1LD\\
$^3$ Harvard-Smithsonian Center for Astrophysics, Cambridge MA 02138\\
$^4$ Department of Physics and Astronomy, University of Southampton, Southampton SO17 1BJ  \\
$^5$ University of Manchester, Jodrell Bank Observatory, Macclesfield, Cheshire SK11 9DL\\
$^6$ Institut d'Astrophysique de Paris, 98bis Boulevard Arago, 75014 Paris\\
}
\date{Accepted ??. Received ??}
\pagerange{\pageref{firstpage}--\pageref{lastpage}}
\pubyear{??}
\maketitle

\begin{abstract}
As part of a multi-wavelength campaign to observe the 2000--2002 outburst of the prototypical symbiotic star Z Andromedae, we observed this object six times each with the Multi-Element Radio Linked Interferometer Network (MERLIN) and Very Large Array (VLA). The radio flux varied significantly during the course of the optical outburst at all three observation frequencies (1.4, 5, and 15 GHz). A jet-like extension was present in the 2001 September MERLIN image and appeared to be aligned perpendicularly to the plane of the binary orbit. Assuming that the ejection took place at the beginning of the optical outburst, the 0.06\arcsec separation between the peak of the extended emission and the central core implies that the ejected material was moving with a velocity of $\sim 400$ km\,s$^{-1}$. This extended emission faded on a timescale of $\sim$ months and was not detected at any other epoch. We consider the implications of jets being a component of a ``prototypical'' symbiotic system and compare properties of the observed jet of Z And with those of the jets in X-ray binary systems.

\end{abstract}


\section{Introduction}

Symbiotic systems contain an evolved red giant star which transfers material onto a more compact hot star, usually a white dwarf (Kenyon 1986). The mass transfer typically proceeds via Bondi-Hoyle capture of the red giant wind and only a small fraction of the wind is accreted. Thus a large nebula forms around the binary and part of the nebula becomes ionized by ultraviolet radiation from the white dwarf.

Every few years, decades or even longer depending on the system, a symbiotic may go into outburst and this picture can change dramatically. The nature and cause of these outbursts, which can have amplitudes of 1--7 magnitudes in the optical and can last from 100 days to many decades, is one of the main outstanding questions for symbiotics. A small subset of symbiotics, the `slow novae' are fairly well established as thermonuclear events that last on the order of decades (Miko{\l}ajewska \& Kenyon 1992). The several symbiotic `recurrent novae', which are much shorter and last on the order of months, are also thought to be thermonuclear runaways, although there has been more controversy on this point (Kenyon \& Garcia 1986; Webbink et al. 1987). The majority of symbiotics (``classical symbiotics''), however, are neither slow novae nor recurrent novae and the cause of their outbursts are not well-understood.

Models for the outbursts of classical symbiotics include expansion of a white dwarf photosphere due to a change in accretion rate onto the white dwarf, thermonuclear shell flashes, accretion disc instabilities or some combination. These outbursts are of particular interest because the white dwarfs in symbiotic stars may accrete enough material to approach the Chandrasekhar mass limit and explode as Type 1a supernovae. A key factor in whether or not symbiotic white dwarfs can approach the Chandrasekhar limit, however, is the amount of material ejected in the outbursts. If more material is ejected from the white dwarf (via e.g. nova explosions or other outflows) than accreted, then clearly the white dwarf mass cannot increase.

\begin{table*}
\caption{MERLIN and VLA flux densities for Z And, derived from two-dimensional Gaussian fits to the images; upper limits are $3\sigma$ values.}
\label{fluxes}
\begin{tabular}{clccccc}
\hline
\hline
JD-2450000 & Observation  & Array&Configuration&  1.4 GHz  &  5 GHz    &  15 GHz  \\
   &  &           &&  mJy (error)  &     mJy (error)   &    mJy (error)\\
\hline
1830& 2000 October 13  &VLA&D   &$<0.29$        &0.42 (0.06)&  0.87 (0.13)\\
1865& 2000 November 17 &VLA&A   &0.28 (0.05)    &0.80 (0.03)&  2.45 (0.09)\\
1921& 2001 January 12  &VLA&A   &$<0.29$        &1.21  (0.10)&   2.94 (0.50)\\
1937& 2001 January 28  &MERLIN&--&   --         &1.32   (0.37)&--\\
1968& 2001 February 28 &MERLIN&--&  --          &0.70   (0.21)&--\\
1983& 2001 March 15    &VLA&B   &$<0.17$	&1.09  (0.03)&  3.50 (0.12)\\
2037& 2001 May 08      &VLA&B   &0.69 (0.02)    &1.02  (0.02)&  2.53 (0.09)\\
2131& 2001 August 10   &VLA&C   &0.28 (0.08)    &1.06  (0.04)&  2.14 (0.14)\\
2170& 2001 September 18&MERLIN&--&   --         &1.20   (0.26)&--\\
2205& 2001 October 23  &MERLIN&--&   --         &0.86   (0.20)&--\\
2394& 2002 April 30    &MERLIN&--&   --         &0.78   (0.20)&--\\
2400& 2002 May 6       &MERLIN&--&    --        &0.64   (0.20)&--\\
\hline
\end{tabular}
\end{table*}


During outbursts the optical spectra of symbiotics often show some evidence for mass-loss in a wind (e.g. Kenyon \& Webbink 1984; Fern\'andez-Castro et al. 1995). In addition, X-ray emission from some symbiotics has been interpreted as due to shock-heated colliding winds from the white dwarf and the mass-donor red giant star (M\"urset et al. 1997). Finally, collimated jets associated with outbursts have been seen directly in the radio maps of several symbiotics -- in particular CH Cyg, the radio emission of which was found to be non-thermal (Crocker et al., 2001) -- and inferred from optical spectra for others (e.g. Hen 1341; Tomov et al. 2000). We discuss this evidence for collimated outflows further in Section 4.

The form of mass-loss can provide some insight into the nature of collimated jets, which are known to be associated with a large variety of astronomical objects from active galactic nuclei (AGN) to proto-stars. It is well-documented that the accretion disc around black hole and neutron star X-ray binaries can produce relativistic ejections (e.g. Fender 2001) and that these objects may sustain more permanent jets (Kaiser, Sunyaev \& Spruit 2000). In this context it is very exciting that white dwarfs with accretion discs are also able to produce bipolar ejections. A study of whether these ejections are more comparable with the low velocity jets from proto-stars or the much faster jets of X-ray binaries is both important and timely.

\subsection{Z And}

Z And is the prototype of the classical symbiotics and is well-known for its dramatic optical outbursts which occur on timescales of a few years to decades (e.g. Formiggini \& Leibowitz 1994). The most recent optical outburst began in 2000 September, reached a peak in 2000 December and had returned to optical quiescence by 2002 August.

The radio counterpart to Z And was discovered by Seaquist, Taylor \& Button (1984) at 4.9 GHz with a flux density of $\sim1.2$ mJy; this observation took place on 1982 February 1 when the source was in a quiescent state. A second set of observations on 1982 July 13 (also during quiescence) at 1.5 and 4.9 GHz yielded a spectral index of 0.62, consistent with a thermal spectrum (e.g. Wright \& Barlow 1975). Similar flux densities were also recorded in 1987 while the source was decaying from the 1984--1986 outburst (Torbett \& Campbell 1989) and during quiescence in 1991--1992 (Kenny 1995). Further observations during the 1984--1986 outburst showed radio variability which appeared correlated with the ultra-violet flux but anti-correlated with the optical (Fern\'andez-Castro et al. 1995); the radio fluxes dropped below the quiescent level by a factor of three during optical outburst.

We have observed Z And since 2000 September as part of a multiwavelength campaign to study classical symbiotic systems in outburst. Observations were obtained at radio, optical, ultraviolet and X-ray wavelengths and the full dataset will be presented in the accompanying paper, hereafter Paper II (Sokoloski et al. in prep.; see also Sokoloski et al. 2002). Here we consider just the radio imaging.

\section{Observations}

\subsection{MERLIN}

High resolution radio images of Z And were obtained at 5 GHz with the Multi-Element Radio Linked Interferometer Network (MERLIN); MERLIN consists of six antennas, of typical diameter $\sim25$m, situated in the U.K. and has a maximum baseline of 217 km. Our observations took place in 2001 January, 2001 February, 2001 September, 2001 October, 2002 April and 2002 May (see Table 1); each epoch included observations of a flux and polarisation calibrator 3C286, a point source calibrator OQ208 and a phase reference source JVAS B2320+506 (alternatively named OHIO Z 533, ICRF J232225.9+505751). The data were reduced using standard flagging, calibrating and imaging techniques within AIPS and the MERLIN ``d-programs''; the telescopes were weighted according to their sensitivity and the data imaged using natural weighting.

\subsection{VLA}

Six observations were also obtained in 2000--2001 by the Very Large Array, in a variety of configurations, at 1.4, 5 and 15 GHz (see Table~\ref{fluxes}). Each epoch included observations of the primary calibrator 3C48 ([VV96] J013741.3+330935) and the secondary calibrator BWE 2352+4933. The data were reduced using standard flagging, calibrating and imaging techniques within AIPS.


\begin{figure}
\begin{center}
\leavevmode  
\psfig{file=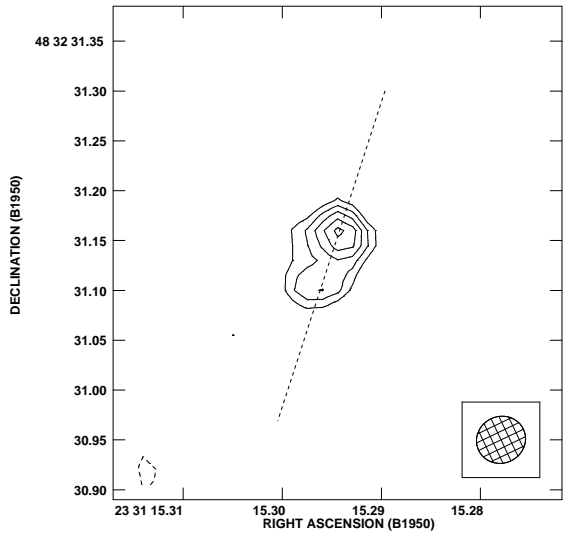,height=7cm, angle=0}  
\caption{MERLIN map of Z And for the 2001 September epoch showing the transient jet-like extension. Contours are plotted for $\sigma\times -3,3,4,5,6,7$. The jet axis is closely aligned with a line drawn perpendicularly to the orientation of the binary orbit as determined by Schmid \& Schild (1997)}
\label{map}
\end{center}
\end{figure}

\section{Results}

\subsection{Radio extension and flux variability}

All images in which a radio source was detected showed point-like emission at the position of Z And, with the exception of the 2001 September MERLIN epoch (at 5 GHz) in which a $\sim60$ mas jet-like extension was resolved to the north-west of the stellar source. We assume that the central core is at the position of peak emission in the unresolved images, which corresponds more closely to the south-east than the north-west point of emission in Fig.~\ref{map}; for comparison the MERLIN spatial resolution is 40 mas and the position error is $\sim10$ mas. The peak flux in the core was $0.27\pm0.05$ mJy/beam compared with a peak of $0.39\pm0.05$ mJy/beam in the extension. Approximate Gaussian fits to the image suggested that the extended emission contained $\sim 55$ percent of the total integrated emission of the source. However, since the peaks of the core and jet were separated by only 40 mas these flux estimates should be treated with caution.

\begin{figure}
\begin{center}
\leavevmode  
\psfig{file=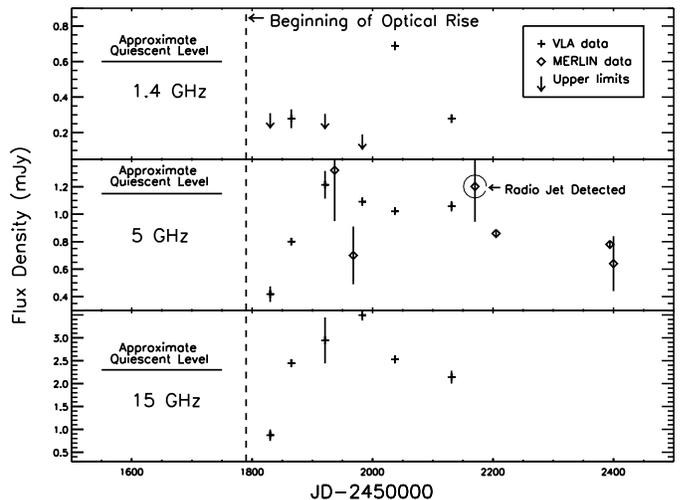,height=7cm, angle=0}  
\caption{Radio lightcurves at 1.4 (top), 5 (middle) and 15 (bottom) GHz. The onset of the optical outburst and the epoch at which the jet was detected are indicated. Typical quiescent values of the flux density are marked by horizontal lines in each plot (e.g. Kenny 1995 and Fern\'andez-Castro et al. 1995 and references therein) and it is clear that the optical outburst was accompanied initially by a decay in the radio emission. The subsequent recovery reached close to the quiescent level at 1.4 and 5 GHz and rose significantly over the quiescent level at 15 GHz (to the highest flux density ever recorded for Z And). }
\label{lightcurve}
\end{center}
\end{figure}

Fig.~\ref{lightcurve} shows the radio lightcurves of Z And at 1.4, 5 and 15 GHz and compares the flux densities during the optical outburst with a typical quiescent level. A more complete multiwavelength analysis will be presented in Paper II but it is clear that the optical outburst was accompanied initially by weak radio flux levels ($\sim 0.4$ mJy at 5 GHz). The radio flux then recovered briefly, reaching 1.4 mJy at 5 GHz and rising above the typical quiescent level at 15 GHz, before decaying again. The peak 15 GHz flux density of $3.5\pm0.12$ mJy is the highest ever reported for Z And. There may also have been a slight flux increase at 5 GHz just prior to our jet detection and a further decay after 2001 September. The 15 GHz VLA fluxes also showed a decline in 2001; this decline was confirmed by supplementary 15 GHz observations taken with the Ryle Telescope, UK in 2002 January (G.G. Pooley, private communication). Values of integrated flux density are listed in Table~\ref{fluxes}. 

An initial drop in flux density below the quiescent level was also observed at 5 GHz in 1986 when it was correlated (but with a possible delay of a few tens of days) with the ultra-violet lightcurve and anti-correlated with the optical (Fern\'andez-Castro et al. 1995). This radio decline of 1986 was interpreted as a reduction in flux of the ionising photons due to the expansion and subsequent cooling of the white dwarf photosphere.

\subsection{A jet in Z And?}

If we define a jet as a collimated outflow, then jets have been discovered in X-ray binaries, active galaxies, young stellar objects and some supersoft X-ray sources. We interpret the extension seen in the 2001 September MERLIN radio image as some form of jet as seen in a number of other symbiotic systems (see Table~\ref{tab:jets}). In the case of the other (unresolved) epochs it is not clear whether the radio emission is from an unresolved jet, the nebula, spherical ejecta associated with the outburst or some combination.

Given the spatial resolution of MERLIN it is possible that the ejection took place at the onset of the 2000--2002 outburst and only became resolvable in 2001 September. Assuming an inclination angle of $\sim 47^{\circ}$ (Schmid \& Schild 1997) and a distance of $\sim 1$kpc (e.g. Kenny 1995), this separation implied that the ejected material was moving at a velocity of $\sim 400$ km\,s$^{-1}$. Analysis of ultra-violet spectra taken by the {\sl FUSE} satellite towards the beginning of the optical outburst suggests absorption by blobs or a patchy shell of material moving with similar velocities  (see Paper II). Since the 2001 October MERLIN epoch revealed a point source, it appears that either this jet was relatively short-lived (it may have expanded and cooled, making it undetectable at 5 GHz) or its emission was absorbed by ejected material associated with the outburst.

Schmid \& Schild (1997) used spectropolarimetry of Z And to measure the inclination ($47\pm12^{\circ}$) and orbit orientation (i.e. the intersection of the orbital plane and the plane of the sky; $72\pm6^{\circ}$). We find that the extended emission in our 2001 September image was consistent with being perpendicular to the orbital plane of the binary, as indicated by the dotted line in Fig.~\ref{map}. This orientation is what we would expect for a binary system in which the compact object is surrounded by an accretion disc; such a disc ejects material perpendicularly as a result of some form of instability, either in the disc or in the mass transfer rate.

Kenny (1995) provides the most detailed study to date of the radio emission from Z And, and compares his images with the STB model (Seaquist, Taylor \& Button 1984; Taylor \& Seaquist 1984). His combined VLA+MERLIN image shows a central band of emission in the plane of the orbit with an additional ``blob'' to the north which may have been ejected during the 1984/5 outburst. We see no significant evidence for a similar central band of emission in any of our images but would certainly expect a contribution to the radio flux from nebular emission in addition to the jet.

\section{Discussion}

\subsection{Possible emission mechanism for the ejected material}

The extended radio emission detected in our observations could arise either from (i) synchrotron radiation produced by relativistic electrons in a magnetic field or (ii) thermal bremsstrahlung of ionised hydrogen gas. Crocker et al. (2001) argue for a significant synchrotron component in the extended radio emission of CH Cyg on the basis of a negative spectral slope. Unfortunately as we do not have resolved images at more than one frequency or sufficient S/N to obtain polarization measurements we can only speculate as to the nature of the radio emission.

We assume that the emission originates in a spherical `blob' of gaseous material, ejected at the time of the start of the optical outburst. To get to its observed position approximately 1 year after the start of the outburst, the blob would have to travel at a constant speed of $v=400$\,km\,s$^{-1}$ (taking the inclination of the jet to be equal to the system inclination $i=47^\circ$; Schmid \& Schild 1997). We estimate the angular size of the blob to be $\theta = 60$\,mas which, for a distance to the source of $d=1$\,kpc, corresponds to a physical diameter of $60 \left( \theta / \theta _{60} \right) \left( d/d_1 \right)$\,AU, where $\theta _{60} = 60$\,mas and $d_1 =1$\,kpc. The volume of the blob is then $4 \times 10^{44} \left(\theta / \theta _{60} \right) ^3 \left( d/d_1 \right) ^3$\,cm$^3$. Rough two-dimensional Gaussian fits to the two components of the source suggest that a fraction of about $f=50$ percent of the total flux density of 1.2\,mJy of the source at 5\,GHz originates in the blob. This flux density corresponds to a monochromatic luminosity of $7 \times 10^{17} \left( d/d_1 \right)^2 \left( f / f_{50} \right)$\,ergs\,s$^{-1}$\,Hz$^{-1}$, where $f_{\rm 50}=50$ percent.

For the radio emission to be (i) synchrotron radiation, the blob must contain a magnetic field and relativistic electrons. As usual, we can only estimate a lower limit for the total energy contained in the field and the electrons using the minimum energy argument (Longair 1994). In the absence of any spectral information, we assume that the energy spectrum of the relativistic electrons follows a power-law with the canonical exponent $p=2.5$. For this choice our results are insensitive to the choice of the upper cut-off frequency of the synchrotron spectrum and we will use $\nu_{\rm max} = 100$\,GHz. For the lower cut-off we adopt $\nu_{\rm min} = 10$\,MHz. Changing $\nu_{\rm min}$ can have a noticeable effect on the derived properties of the blob, but our results are only order-of-magnitude in any case and so even changes by factors of order 10 do not change our conclusions. Using these assumptions we find a minimum total energy stored in the magnetic field and the relativistic particles in the blob of 

\begin{equation}
E_{{\rm int} i} = 2 \times 10^{39} \left( \frac{d}{d_1} \right) ^{\frac{17}{7}} \left( \frac{\theta}{\theta _{60}} \right) ^{\frac{9}{7}} \left(\frac{f}{f_{50}} \right)^{\frac{4}{7}} \,{\rm ergs.}
\end{equation}

\noindent
The magnetic field strength for this case would be $8\left( d/d_1 \right) ^{-2/7} \left( \theta / \theta _{60} \right)^{-6/7} \left( f/f_{50} \right) ^{2/7}$\,mG and the total number density of relativistic electrons is 

\begin{equation}
n_{e,i} = 5 \times 10^{-2} \left( \frac{d}{d_1} \right) ^{-\frac{5}{7}} \left( \frac{\theta}{\theta _{60}} \right) ^{-\frac{15}{7}} \left(\frac{f}{f_{50}} \right) ^{\frac{5}{7}}$\,cm$^{-3}.
\end{equation}

\noindent
Under these conditions, the optical depth of the blob due to synchrotron self-absorption is $2\times 10^{-8} \left( \theta / \theta _{60} \right) ^{-2} \left( f/f_{50} \right)$ and so absorption is negligible. If the electrostatic charge of the relativistic electrons is balanced by one proton for each electron, then the bulk kinetic energy of the blob is 

\begin{equation}
E_{{\rm kin},i}=2\times 10^{34}\left(\frac{d}{d_1}\right)^{\frac{16}{7}}\left(\frac{\theta}{\theta_{60}}\right)^{\frac{6}{7}}\left(\frac{f}{f_{50}}\right)^{\frac{5}{7}}\left(\frac{v}{v_{400}}\right)^2{\rm ergs}
\end{equation}

\noindent
where $v_{400} = 400$\,km\,s$^{-1}$.

In the case of (ii) bremsstrahlung emission, we assume that the blob is filled with pure, ionised hydrogen gas at a temperature of $10^4$\,K. Using the expressions for thermal bremsstrahlung emissivity in Longair (1994) we require a proton density within the blob of 

\begin{equation}
n_{p,ii} = 7 \times 10^5 \left( \frac{d}{d_1} \right)^{-\frac{1}{2}} \left( \frac{\theta}{\theta_{60}} \right) ^{-\frac{3}{2}} \left( \frac{f}{f_{50}} \right) ^{\frac{1}{2}} \left( \frac{T}{T_4} \right)^{\frac{1}{4}}{\rm cm^{-3}}
\end{equation}

\noindent
where $T_4 = 10^4$\,K. Here we neglect the extremely weak temperature dependence of the Gaunt factor and the exponential part of the relevant expression. This density corresponds to a total mass of 

\begin{equation}
M_{{\rm blob},ii} = 2\times 10^{-7}\left(\frac{d}{d_1}\right)^{\frac{5}{2}}\left(\frac{\theta}{\theta_{60}}\right)^{\frac{3}{2}}\left(\frac{f}{f_{50}}\right)^{\frac{1}{2}}\left(\frac{T}{T_4}\right)^{\frac{1}{4}}M_{\odot}
\end{equation}

\noindent
for the blob. The total thermal energy of the blob material is

\begin{equation}
E_{{\rm int},ii} = 6 \times 10^{38}\left( \frac{d}{d_1} \right) ^{\frac{5}{2}} \left(\frac{\theta}{\theta_{60}} \right) ^{\frac{3}{2}} \left( \frac{f}{f_{50}} \right) ^{\frac{1}{2}} \left(\frac{T}{T_4} \right)^{\frac{5}{4}}\,{\rm ergs}
\end{equation}

\noindent
while its bulk kinetic energy is

\begin{equation}
E_{{\rm kin},ii} = 4 \times 10^{41}\left(\frac{d}{d_1} \right) ^{\frac{5}{2}} \left(\frac{\theta}{\theta_{60}} \right) ^{\frac{3}{2}} \left(\frac{f}{f_{50}} \right)^{\frac{1}{2}} \left(\frac{T}{T_4} \right)^{\frac{1}{4}} \left( \frac{v}{v_{400}}\right)^2 {\rm ergs.}
\end{equation}

\begin{table*}
\begin{center}
\caption{Symbiotic stars and supersoft X-ray binaries showing evidence for jets\label{tab:jets}} 
\begin{tabular}{cccccc}
\hline
\hline
Source Name & Instrument & Velocity (km s$^{-1}$)&  P$_{orb} (d)$ & Comments & Ref.\\
\hline
$\underline{\rm Symbiotic\; Stars}$ & & & & & \\
RS Oph & EVN &  3800  & 455.7 & outburst (RN), non-thermal & 21\\
CH Cyg & VLA, HST & $\sim$ 600--2500 & $\sim$5700 &post-outburst, non-thermal & 8, 10, 13, 18, 20\\
Hen 1341 & opt spec &  $\sim$ 820  & & outburst & 22\\
MWC 560 & opt spec & $\sim$ 600-2500 & 1930 & discrete ejections & 17, 23\\
StH$\alpha$ 190 & opt spec & $>$ 1000  & $\sim$170? & no outbursts & 15\\
R Aqr & HST, Chandra, MERLIN & $\ge 300$& $\sim$ 16000 & shock-heated gas, precession? & 4, 9, 11\\
Z And & MERLIN & $\sim$ 400 & 758.8 & outburst & 2\\ 
AG Dra & MERLIN & 800& 554 & post-outburst, thermal & 12, 16 \\
HD 149427 & HST (possibly), ATCA &&&discrete ejections?&3\\
V1329 Cyg & HST&$\sim$ 250&956.5&Flares?&3\\
 & & & & & \\
$\underline{\rm Supersoft\; Binaries}$ & & & & & \\
RX J0925.7-4758 & opt spec & $\sim$5000 & 4.0 & transient jet & 14\\
RX J0513-69 & opt spec & $\sim$4000  & 0.76 & persistent jet & 5, 7, 19\\
RX J0019.8+2156 & opt spec & 3500 - 4000  & 0.66 & jet precession & 1, 5, 24\\
CAL 83 & opt spec & $\sim$700 - 2500 & 1.0& jet precession& 5, 6\\
\hline 
\end{tabular}
\begin{minipage}{160mm}
\vspace{0.3cm}
REFERENCES -- 
1. Becker et al. 1998,
2. Brocksopp et al. 2003 (this work),
3. Brocksopp, Bode \& Eyres 2003,
4. Burgarella \& Paresce 1992,
5. Cowley et al. 1998,
6. Crampton et al. 1987,
7. Crampton et al. 1996,
8. Crocker et al. 2001,
9. Dougherty et al. 1995,
10. Karovska et al. 1998,
11. Kellogg et al. 2001,
12. Miko{\l}ajewska 2002,
13. Miko{\l}ajewski \& Tomov 1986,
14. Motch 1998,
15. Munari et al. 2001,
16. Ogley et al. 2002,
17. Schmid et al. 2001,
18. Sokoloski \& Kenyon 2003a,
19. Southwell et al. 1996,
20. Taylor, Seaquist \& Mattei 1986,
21. Taylor et al. 1988,
22. Tomov, Munari, \& Marrese 2000,
23. Tomov et al. 1990,
24. Tomov et al. 1998

\end{minipage}
\end{center}
\end{table*}

\noindent
The optical depth to bremsstrahlung absorption of the blob is roughly $0.9 \left( \theta / \theta _{60} \right) ^{-2} \left( f /f_{50} \right)$ which implies that the blob is almost opaque. Finally, at the measured emission rate, the thermal energy of the blob material would be radiated away within about $200 \left(d/d_1 \right)^{7/2} \left( \theta / \theta _{60} \right) ^{9/2} \left(f/f_{50} \right) ^{-1/2} \left( T/T_4 \right) ^{1/4}$\,days.

The (i) synchrotron scenario could explain the observations but also has potential problems. The optical depth of the blob is extremely small and so if the blob was ejected at the beginning of the optical outburst, then perhaps it should have been detected earlier. Alternatively the early non-detection of the blob could be explained by the initially unresolvable separation between the blob and the radio core, and also by the drop in nebular emission early in the outburst, which could have dominated over any rise in radio flux due to the jet. Because of the long radiative lifetime of the relativistic electrons of about $800 \left( d/d_1 \right) ^{3/7} \left( \theta / \theta _{60} \right) ^{9/7} \left( f/f_{50} \right) ^{-3/7}$ \,yr under the derived conditions in the blob, the jet might also have been expected to have been detected at later epochs. However, continued adiabatic expansion of the blob could lead to a faster dimming than implied by pure radiative energy losses. Finally, and most problematically, the inferred density of the relativistic particles is very small. This low density implies that the blob would have to contain a very significant amount of thermal material as well as  relativistic gas in order to push its way out through the dense, inner regions around the white dwarf (e.g. Mikolajewska \& Kenyon 1996) at the inferred velocity of 400\,km\,s$^{-1}$. However, if the blob material is dominated by thermal material, then, unless the material is very cold, scenario (i) will resemble scenario (ii).

For (ii) bremsstrahlung emission the situation looks more promising. Even allowing for some expansion of the blob on its way out from the source centre, the inferred density of the blob material is probably high enough for the blob to push through the dense inner regions of the system. Furthermore the optical depth would indicate that the blob emission could have been completely absorbed before the time of detection and the short cooling time of the gas would explain why we did not observe the blob at later epochs.  A relatively large bulk kinetic energy of the blob is required (see Equation 7) but only a small fraction of the envelope mass ($\sim$ a few times $10^{-5}$\,M$_{\odot}$; see Fujimoto 1982) would need to be burned to produce this amount of energy and so this should not pose a problem to the bremsstrahlung model.  The integrated luminosities of novae are $\sim 10^{46}$ erg and the kinetic energy of the ejected material is thought to be similar (Warner 1995). However, it is not clear whether classical symbiotic outbursts involve an increase in the thermonuclear burning rate.  If the outburst was due entirely to an accretion instability, accretion at $6 \times 10^{-8}$ M$_{\odot}$ yr$^{-1}$ for 2 weeks could provide $\sim 10^{41}$ erg.  Accretion rates in symbiotics are thought to be a few times $10^{-8}$ M$_{\odot}$ yr$^{-1}$ on average, so an increase to around $10^{-7}$ M$_{\odot}$ yr$^{-1}$ during outburst is plausible.  Alternatively, the maximum estimated blob energy, $4 \times 10^{41}$ erg, could be produced in just (for example) 50 days at a luminosity of 0.25L$_{\odot}$. M\"urset et al. (1991) derive hot component quiescent luminosities of $>600$L$_{\odot}$ and so we cannot rule out a significant thermonuclear contribution. Scenario (ii) therefore places interesting constraints on outburst energetics.

A possible alternative could be based on a more continuous outflow rather than a discrete blob ejection. In this case, the emission may be bremsstrahlung from the compressed and heated gas pushed aside in front of the outflow. The presence of an outflow would not significantly relax the energy requirements for the emission region, but the energy would be delivered over a longer time rather than instantaneously. There is some observational evidence that outflows may be a common feature of symbiotics even in quiescence.  Skopal et al. (2002) have done comprehensive studies of outflows in CH Cyg, and M{\"u}rset, Wolff, \& Jordan (1997) suggested that colliding winds from the red giant and accreting white dwarf produce the X-ray emission that is seen in roughly 60 percent of the symbiotics in the ROSAT database. However, the degree of collimation of these outflows is not known.

\subsection{Comparison with other jet-emitting symbiotic systems}

The presence of a jet in a ``prototypical'' symbiotic system such as Z And has important implications for symbiotic star research. While jets have been imaged at radio frequencies in e.g. CH Cyg (Crocker et al. 2001) and R Aqr (Dougherty et al. 1995), these sources are atypical and jets have not been considered to be standard features of a symbiotic system. However an increasing number of other symbiotic systems and also some of the supersoft binary sources are now thought to have some sort of jet or bipolar outflow on account of radio/optical imaging or radial velocity measurements. These sources, along with derived jet velocities, are listed in Table~\ref{tab:jets}.

Interestingly, the sources listed do not appear to have any other readily apparent properties in common with one another, besides being symbiotic stars or supersoft binaries (i.e., white-dwarf accretors that probably have higher accretion rates than cataclysmic variables). The outflow velocities range from a few hundred to a few thousand km\,s$^{-1}$ and some, {\em but not all}, of the sources were in outburst at the time of ejection. The symbiotic type varies with most being S-type, but R Aqr being D-type and StH$\alpha$, HD 149427 being D$\arcmin$-type. Quasi-steady nuclear burning is thought to take place on the surface of Z And, AG Dra and the supersoft sources, but not in the case of RS Oph, CH Cyg and R Aqr (e.g. van den Heuvel et al. 1992, Dobrzycka et al. 1996, Sokoloski et al. 2001, M\"urset et al. 1991). Z And is the only one thought to have a highly magnetised white dwarf (Sokoloski \& Bildsten 1999; but see also Tomov et al. 1996, Tomov 2003, Sokoloski \& Kenyon 2003b). Therefore, other than the presence of the white dwarf, it is difficult to find any property common to these systems which is of relevance to the jet. The presence of an accretion disc is an obvious possibility, but direct evidence for (or against) the existence of an accretion disc is difficult to find for symbiotic stars\footnote{Cataclysmic-variable-like rapid optical variability (as seen in e.g. RS Oph, CH Cyg and MWC 560) is one potential disc indicator.  See Sokoloski (2003) for a discussion of optical flickering as a diagnostic of accretion in symbiotics.}.

\subsection{Comparison with X-ray binaries}

In order for the extended radio emission from Z And to be a radio jet comparable with those seen in other types of system, we would expect to see a number of comparable properties. All other jet systems show evidence for an accretion disc and this is one of the requirements of jet models. While no direct evidence has yet been found for an accretion disc in Z And, Sokoloski \& Bildsten (1999) detected a 28-minute coherent oscillation in the optical lightcurve. This modulated light was interpreted as emission from the accreting polar caps of a rotating and  magnetic white dwarf, leading to difficulties in explaining the small outburst of 1997--1998 in terms of thermonuclear runaway on the white dwarf surface since the oscillation was present {\em during} the outburst; an accretion disc instability is a feasible alternative. 

Fender \& Hendry (2000) found that neutron stars with {\em weak} magnetic fields were more likely to have synchrotron jets than those with strong ($10^9-10^{10}$ G) magnetic fields, perhaps due to disruption of the accretion flow caused by the latter. Z And, however, is a magnetic system (Sokoloski \& Bildsten 1999) and although the magnetic field of Z And is lower than that of X-ray pulsars, it is sufficient to produce a comparable degree of truncation of the inner disc. Therefore the presence of a jet in Z And is consistent with there being a different formation mechanism for thermal and non-thermal jets.

An additional property observed in black hole X-ray binaries is that while a jet is present during the low/hard state, it becomes ``quenched'' when the system makes a transition to the high/soft state (e.g. Fender 2001). This quenching is apparently due to changes in the geometry of the accretion disc and/or mass accretion rate, although the exact nature of the jet:disc relationship is still undetermined. Similarly, the apparent reduction in radio flux in Z And early in the optical outburst could be an important property if it is intrinsic to the jet rather than due to a decrease in size of the ionized nebula (through the expansion and subsequent cooling of the white dwarf photosphere; see Fern\'andez-Castro et al. 1995). Conversely outflows from FU Ori stars increase during the high state of the unstable disc, and so the detection of extended radio emission from Z And during outburst could be more consistent with this picture (Pudritz \& Ouyed 1999).

The jets of black hole X-ray binaries in the low/hard state have a flat synchrotron radio spectrum (Fender 2001) which is intrinsic to the jet itself and not the result of shocked emission due to interaction with nebula material, which was the case in CH Cyg (Crocker et al. 2002). Spectral indices measured for Z And from the 5 and 15 GHz VLA data lie in the range 0.6--1.0, consistent with thermal radio emission (e.g. Wright \& Barlow 1975). We note however, that none of these spectral index measurements was obtained contemporaneously with the jet and so we cannot rule out a non-thermal contribution.

The ejection of blobs of material from Z And in 2001 (this work) and 1991 (Kenny 1995) both appear to have been one-sided, with the ejection at a similar posotion angle relative to the central source in each epoch. However, since the image is only marginally resolved it is still possible that the jet is indeed double-sided. Images of e.g. CH Cyg have also appeared one-sided (see Table~\ref{tab:jets} for references) and so one-sidedness may be a genuine property of some jet-emitting symbiotic systems. While one-sided jets are not uncommon in active galaxies and X-ray binaries, they are usually explained in terms of Doppler boosting effects due to the relativistic speeds of the jet. Since there is unlikely to be relativistic motion taking place in a symbiotic system we suggest that the most likely explanation is absorption of the emisison of the second jet by the red giant wind or some form of patchy shell. The presence of such a shell is implied by our ultraviolet spectra and will be discussed in detail in Paper II. Alternatively the second jet in the Z And system may have been retarded sufficiently by material ejected during the outburst for it to be unresolvable (again, we address this further in Paper II in conjunction with our ultra-violet observations). 

\section{Conclusions}

Radio imaging of Z And with MERLIN over the course of the recent outburst has resolved a jet-like extension perpendicular to the orbital orientation during one epoch. Monitoring with the VLA and MERLIN during the 2000--2002 outburst revealed that initially the radio flux density dropped below typical quiescent levels; it then recovered and later faded again. Whether this drop was due to increased absorption, an intrinsic property of the radio jet or a decrease in the size of the ionized nebula is not yet certain. We have compared Z And with jet-emitting X-ray binaries and find that there is a greater number of important differences than similarities; thus while the potential implications of a jet in such a prototypical system are highly significant we cannot consider the jet an analogue of an X-ray binary jet. Current data show more similarities with the outflows in supersoft X-ray sources.  Furthermore, a comparison of Z And with other symbiotic jet sources reveals that there is no obvious unifying observable property for symbiotic jet sources, although they could all contain accretion discs.  Whether symbiotic jets are more like X-ray binary jets or the slower supersoft X-ray source and young stellar object jets, they promise to provide useful information about jet production.

\section*{acknowledgements}
We thank Scott Kenyon for useful comments. MERLIN is a UK National Facility operated by the University of Manchester on behalf of PPARC. JS acknowledges support from NSF grant INT-9902665 and NASA grant NAG5-11207.

\end{document}